\documentclass[a4paper,11pt]{article}

\usepackage[breaklinks=true]{hyperref}
\usepackage{breakcites}

\usepackage{amssymb,amsmath,array,bm}
\usepackage{booktabs}

\usepackage{tikz}
\usetikzlibrary{arrows}

\usepackage[a4paper, left=3cm, right=2.5cm, top=3cm, bottom=3cm]{geometry}

\usepackage{authblk}
\usepackage{natbib}
\usepackage{doi}

\usepackage{amssymb}

\usepackage[utf8]{inputenc}
\usepackage[T1]{fontenc}
\usepackage[english]{babel}
\usepackage{csquotes}

\title{Healthy distrust in AI systems}

\author[1]{Benjamin Paaßen}
\author[2]{Suzana Alpsancar}
\author[2]{Tobias Matzner}
\author[2]{Ingrid Scharlau}
\affil[1]{Bielefeld University, Germany}
\affil[2]{Paderborn University, Germany}

\date{Preprint as provided by the authors.}

\begin{document}

\maketitle

\pagestyle{myheadings}
\markright{Healthy distrust. Preprint as provided by the authors.}

\begin{abstract}
Under the slogan of trustworthy AI, much of contemporary AI research is focused on designing AI systems and usage practices that inspire human trust and, thus, enhance adoption of AI systems. However, a person affected by an AI system may not be convinced by AI system design alone---neither should they, if the AI system is embedded in a social context that gives good reason to believe that it is used in tension with a person’s interest. In such cases,  distrust in the system may be justified and necessary to build meaningful trust in the first place. We propose the term \emph{healthy distrust} to describe such a justified, careful stance towards certain AI usage practices. We investigate prior notions of trust and distrust in computer science, sociology, history, psychology, and philosophy, outline a remaining gap that healthy distrust might fill and conceptualize healthy distrust as a crucial part for AI usage that respects human autonomy.
\end{abstract}

\section{Introduction}

Making AI systems trustworthy is a central pursuit of current AI research, with entire research fields devoted to enhancing fairness, explainability, robustness, and further system properties that, together, form a notion of trustworthy AI \citep{EU2019,Buschmeier2023,Ji2024,Kaur2022,Huang2024}. These efforts have also extended into policy, e.g.\ with the aim of codifying trustworthy AI in the European AI act \citep{Laux2024}. However, the quest for ever more trustworthy AI systems leads away from one fundamental question: Aren't there many cases in which \emph{distrust} is desirable?

We argue that there are indeed scenarios where distrust is \emph{healthy}, namely when human autonomy and justified interests are in tension with reliance on an automatic system. In such scenarios, we argue, a notion of healthy distrust is needed to a) accurately describe the empiric reality that some humans have good reason to distrust a system even though it checks all boxes of a trustworthy AI system, and b) recognize that distrust can be a desirable state in human-AI-interaction that ought to be fostered and cultivated in order to improve the relationship of autonomous citizens with AI systems. Accordingly, our proposed term, \emph{healthy distrust}, emphasizes both descriptive and normative aspects.

We acknowledge that distrust, as such, has already been extensively researched. The pursuit for trustworthy AI is grounded in the recognition that unfairness, lack of transparency, etc.\ may be good reasons for distrust which ought to be addressed \citep{Kaur2022}. Psychology has developed scales to measure trust in AI systems and investigated conditions under which trust is lacking \citep[e.g., ][]{hoffman2023measures, Koerber2018trust, scharowski2024trust}. Further studies have investigated automation bias, i.e.\ the human tendency to over-rely on automatic systems, e.g.\ under time-pressure \citep{Goddard2011}.

Still, these prior works tend to conceptualize trust and distrust as opposing notions on a single scale, whereas recent research suggests that, instead, trust and distrust towards the same system can co-exist \citep{peters2025fip,schul2008JEPS}. Indeed, this paradoxical co-existence may be typical for healthy distrust: When we accurately understand the possibilities and limitations of an AI system, it is appropriate to trust the system to fulfill tasks it was trained for but to \emph{distrust} the system to generalize beyond this scope \citep{Ilievski2024}. Similarly, healthy distrust can enrich our understanding of explanations in AI. While explanations are often understood as measures to promote trust, an accurate explanation may also provide reasons to \emph{distrust} the system in some respects---which may be an appropriate, healthy reaction.

Overall, this work is an attempt to engage in interdisciplinary cooperation to uncover gaps in disciplinary understandings of trust and distrust, and to jointly move forward towards a notion of healthy distrust that can push interdisciplinary AI research (and practice) forward.

\section{Going beyond trustworthy AI}

The most common notion of trust used (implicitly) in the technical disciplines is that we (should) trust a system if it reliably fulfills the purpose it was designed for as certified by a technical evaluation study \citep{Birhane2022}. Indeed, improving quantified measures of technical performance is the main pursuit of much (if not most) AI research: AI researchers aim to develop methods and systems that reliably generalize a function to new data \citep{Ilievski2024}. However, while technical reliability and performance is certainly one aspect of trustworthiness of AI systems \citep{Kaur2022,EU2019}, many AI researchers have recognized the shortcomings of such a narrow focus. For one, it has become increasingly obvious that more general-purpose AI systems, such as large language models, are used far beyond the purpose they were trained for, such that any traditional notion of generalization and reliability breaks down \citep{Ilievski2024}. Even more importantly, though, more widespread use has made it increasingly obvious that there are good reasons to distrust systems \emph{even if} they fulfill their specified technical purpose. The most prominent examples are cases of algorithmic discrimination, e.g.\ different performance across gender and skin color for face recognition systems \citep{Buolamwini2018}, or unjustified high risk scores for Black people in the US criminal justice system \citep{Angwin2016}. Since then, regulators and researchers have come up with more and more extensive lists of requirements for \emph{trustworthy AI} systems \citep{Kaur2022}. For example, the EU guidelines for trustworthy AI define \enquote{Human agency and oversight}, \enquote{Technical robustness and safety}, \enquote{Privacy and data governance}, \enquote{Transparency}, \enquote{Diversity, non-discrimination and fairness}, \enquote{Societal and environmental wellbeing}, and \enquote{Accountability} as central requirements \citep{EU2019}. The survey by \citet{Kaur2022} states that the most common requirements listed are \enquote{Fairness}, \enquote{Explainability}, \enquote{Accountability}, and \enquote{Privacy}. Similarly, much AI research has aimed to develop methods that \emph{align} AI systems with human values, focusing on Robustness, Interpretability, Controllability, and Ethicality \citep{Ji2024}. Such frameworks for trustworthy AI have substantially broadened the scope and complexity of considerations when researching AI systems and applying them in practice -- and they have formed an important basis for AI regulation, at least in the EU context \citep{Smuha2021}. Based on these frameworks, researchers have developed measures to assess trustworthiness \citep{schlicker2025we} in systems and engineer systems for trustworthiness \citep{kaestner_etal2021trust}. However, we note that such frameworks still focus on \emph{system requirements} in the sense that they set up demands on how systems ought to be designed, trained, and used to be worthy of trust. The implicit assumption appears to be: once this checklist of system requirements is crossed off, we expect people to trust the systems in question (and become enthusiastic users and customers). If this is indeed the view implicitly held, at least two questions are ignored: The empiric psychological question whether humans are actually likely to trust systems if they comply with the requirements listed; and the normative question of whether humans \emph{should} trust systems if they comply with the requirements listed.

To illustrate the complexities, consider a scoring system that aims to estimate a person’s future risk for a wide range of possible illnesses. If such a system is employed by a health insurance company to increase the insurance premium for a person (or decline coverage altogether), said person would likely distrust the system, even if it had been certified by independent agencies according to all kinds of requirements and regulations. In particular, we might argue that the person is justified in their distrust, e.g.\ because the company has a monetary incentive to overestimate risks and, therefore, overcharge their customers (even if they empirically do not); or because the user does not have sufficient insight into the inner workings of the scoring mechanism; or because such high stakes decisions should never be made by a machine in the first place \citep{Barocas2023}. 


More broadly, we are able to come up with reasons for distrust that go beyond the lack of compliance with a list of design or usage requirements and that technological system design can not address by itself. Indeed, healthy distrust may call into question whether automation in a particular scenario is desirable in the first place. Accordingly, we have to turn to other disciplines and broaden our focus, both to understand healthy distrust better, and to find ways to foster healthy distrust. In the following sections, we cast our nets into the seas of history, sociology, psychology and philosophy.

\section{A short history of trust}
To contemporary readers, it may seem obvious that trust is crucial when interacting with AI. From a historical perspective, this is not self-evident. \citet[p. 174]{kaminski2019Begriffe} terms trust a \enquote{central marginal note in Western thinking}, referring to the observation that for several centuries, it played a minor role and its conceptualization was rather scattered. German historian Ute Frevert has argued that, in modern societies (she refers mostly to the 18th, 19th, and 20th centuries), questions of trust are everywhere\footnote{Her analysis is restricted to German documents. There is little reason to suspect large differences to other Western countries which is why we use the analysis here as a general frame.}. What is more, trust usually comes an appeal or plea to individuals. She calls trust an \enquote{obsession of modernity} \citep{frevert2013Vertrauensfragen}. Frevert’s analysis focuses trust; distrust is much less in focus, but appears as the opposite of trust, which is, as we will see, a common, but not necessary figure.

During the period analyzed by Frevert, what the term \enquote{trust} refers to has shifted and, in particular, expanded. While in the 18th century, it was largely limited to trust in God, in the 19th century, the meaning shifted towards family and friend relationships and gained more affective meaning as well as reciprocity. It was strongly valorized as a moral resource. \enquote{Modern humans showed themselves as fundamentally trusting and trustworthy; however, they must not be too trusting either} \cite[][p. 213, own translation]{frevert2013Vertrauensfragen}. On a more explanatory level, Frevert sees the wish for personal trust as a side effect of individualization, which in turn was an effect of the development of modern civil societies. 

In the 20th century, trust was increasingly extended to organizations (such as trade and craft) as well as functionaries (such as employers or politicians). As it became \enquote{increasingly linked to functions, competencies and objectifiable achievements} \cite[][p. 216, own translation]{frevert2013Vertrauensfragen}, the term lost some of the moral aspect it had had in the 19th and 18th centuries. In the second half of the 20th century, Frevert observes that, surprisingly, trust is requested in all functional relationships between people and, thus, the term turns into a primarily ethical/moral concept.

Two developments on the periphery of the development observed by Frevert are worth mentioning. Firstly, in economics, the increasingly prominent rational choice theories assign trust an important function, but see it as a calculation and, thus, as an expectation of the future; thereby, it loses its important emotional quality according to \citet[][p. 16]{frevert2013Vertrauensfragen}. This emotional quality is also evident in the consistently positive connotation of trust in the period she analyzed. Secondly, a more naturalized understanding appears, most prominently in psychoanalyst Erik H.\ Erikson’s notion of a basic trust that is necessary for healthy development and a healthy personality. \enquote{This lent the term a high, almost unassailable moral value, which also seemed to have a biological-natural foundation} \cite[][p.\ 40, own translation]{frevert2013Vertrauensfragen}. In the late 20th and early 21st century, the ability to trust, thus, became, across spheres, a measure of stable mental health. Trust entered into objectified relationships to which it lends a special quality. 

The perhaps strongest expansion concerns the new contexts of consumption and services and an increased relevance in politics in the second half of the 20th century. In public discourse, trust almost completely replaced alternatives such as confidence \cite[relating to Luhmann’s notion of system trust][]{luhmann2000vertrauen}.

One strong conclusion of Frevert’s is: \enquote{To evoke trust, therefore, means to work on the humanization, emotionalization and personalization of the impersonal---and to profit from it} \cite[][p. 218, own translation]{frevert2013Vertrauensfragen}. In this perspective, mistrust or distrust has a strongly negative connotation---as the opposite of trust, it lacks this positive emotional and personal quality.

Mistrust has been much less studied and has been in the focus only in the last two decades. If studied, it \enquote{is usually treated as the flip side of trust, as an annoying absence, a societal failure, or an obstacle to be overcome} \cite[][]{mühlfried2018intro}. Mühlfried, however, points out that mistrust also serves psychological, societal, and epistemological functions. As trust, it can reduce complexity. It can also mean to engage with complexity if people question and try to understand seemingly given things---for instance, in the form of doubt and scrutiny.

\section{Conceptualizing trust in practical philosophy}
From the point of view of practical philosophy, trust is of interest as a notion associated with our capacities of acting. To trust someone or something allows us to act in a certain way, which we otherwise could not. Trust broadens our scope of action under conditions of uncertainty. We may differentiate between trust in a general and in a specific sense. Trust in general is as pervasive as it is necessary; nothing would function or work out if most people wouldn't constantly trust most other people and things. 

\begin{quote}
Imagine a world without trust. We would never enter a taxi without trusting the driver’s intention and the car’s ability to bring us to our desired destination. We would never pay a bill without trusting the biller and the whole social and institutional structures that have evolved around the concept of money. We would never take a prescribed drug without any trust in the judgment of our doctor, the companies producing the drug, and all the systems of care and control that characterize our healthcare system. We would not know when and where we were born if we would distrust the testimony of our parents or the authenticity of the official records.\citep[p. 1]{simon2020routledge}
\end{quote}

However, this general trust is not of interest for the question of healthy distrust in AI as this concerns a more specific relation: \textit{When is it warranted to trust the AI in doing $\Phi$ or to $\Phi$, and when not?} And \textit{What does it mean to possess healthy distrust against an AI system's performance in general, as well as in regard of a particular instance?} Likewise, we may rule out of general distrust, i.e., radical scepticism, as it is practically just not feasible to distrust everything all at once, or even to distrust very many things all at once. In terms of more specific (dis-)trust, the theoretical tradition in practical philosophy and social science offers some fruitful distinctions and conceptual clarifications that we draw on, here. In particular, we aim to shed light onto the question of 
what is particular about (dis-)trusting AI systems in comparison with other trust relationships. 

20th century trust theories offer two paradigms of trust relationships: trust between two persons, and trust as a necessary condition for cooperation. The first one derives from everyday experiences in friendship and family live, the second reflects everyday experiences in modern societies and its typical division of labor and systemic, functional differentiation. As we will see, the relationship to AI partly resembles our relationship with other persons and partly our relationship to social systems, structural and organizational processes and conditions, etc. 


Of particular interest for our question are the philosophical debates on trust and reliance, the different perspectives on trust proposed by the the predicative vs.\ the affective view of trust, as well as the link between trust and risk. 

First, there is a debate about whether or not trust equals reliance \citep{goldberg2020trust} or if it accounts for something more \citep{lahno2020trust}. Considering this distinction is of interest, because there is a broad consensus that people rely as much on artifacts as they do on other persons, while there is no such consensus in terms of trust. Some argue that speaking of trust in artifacts, artificial agents, is a category mistake, because \enquote{genuine trust} is something we may only put in other persons \citep{freiman2023making}. The reasons usually given for drawing such a distinction is that technical artifacts are not part of our shared social and moral praxis which is grounded in (the possibility of) mutual recognition. Genuinely trusting another person implies to be vulnerable as a person oneself, e.g., in form of the possibility of getting betrayed \citep{baier1986trust}: If the trustee lets the trustor down, this will not only harm whatever function the object of trust was meant to fulfill, but also negatively affect the relationship between the two persons. An indication for this are typical morally and socially loaded responses to trust violation, such as feeling betrayed, abused, not recognized as a person, or not valued in our human dignity. There are also positive emotions towards trust fulfilling, such as gratitude. Applying such a strong concept of genuine trust on machines does appear nonsensical: What should it mean to feel betrayed by your car?

While the idea of \enquote{genuine trust} implies to ascribe human qualities on the trustee, such as goodwill and moral obligation, \enquote{mere reliance is a way of acting in light of the probability that technology will perform successfully} \citep[p. 1353]{freiman2023making}. Hence, speaking of reliance in AI would not evoke these kinds of conceptual awkwardness. If we follow that distinction, we might reconsider what the \enquote{AI and trust} debate is actually about. While trustworthy AI is regularly linked to reliance (and reliability), our impression is that this link is not well-defined and that it is not so clear what \enquote{trusting} AI is meant to mean beyond mere reliance. Critics argue that talk of trustworthy AI is a marketing ploy, a charade put on by industry and politicians to win over as many AI consumers as quickly as possible \citep[for a nuanced discussion see][]{freiman2023making}. However, we may conclude that speaking of trustworthy AI adds to the pervasive anthropomorphization of AI which deeply affects our understanding of what AI is and can do, which might be problematic as such \citep{Hasan2024-HASWYA-2, deshpande2023anthropomorphizationaiopportunitiesrisks,Troshani2021trust}.  

Obviously, this differentiation between mere reliance (or trust as reliance) and genuine trust raises the question in how far we perceive AI systems to be human-like. Most argue that AI systems are more than simple technical artifacts (or complex machines), and, hence, trusting an AI (e.g., to give appropriate advice on medical interventions) could mean something more than relying on a boat to stay afloat \citep[p. 301]{grodzinsky2020trust}.

The tendency and extension to anthropomorphize some AI-based systems seems to be much higher in contrast to the also common anthropomorphization of other (mundane) technical artifacts such as toys, cars, etc. Although people hold affective relations to their cars---speak to them and don't want them to become dysfunctional, etc.---this emotional attachment seems to be grounded in acknowledging the difference between the technical artifact and a person. For instance, even car enthusiasts are not known for having tried to marry them. This distinction in terms of our social practices seems to get increasingly blurred with some AI systems, such as conversational AI. While we can speak to cars and communicate with pets, conversational AI speaks our language. Here, it would be worth to further explore, how the idea of \textit{healthy distrust} relates to the tendency to anthropomorphize AI: does it (or should it) counterbalance this tendency? Would healthy distrust also include some sort of skepticism-disposition towards AI development and deployment? 

Second, another dominant debate in practical philosophy and social science is about whether viewing trust as something a rational agent might actively choose, or viewing trust as something that is not reducible to rational decision-making because it is affective, emotional, and something pre-given to deliberative actions. The first account is often called \enquote{predictive view} or \enquote{rational view}, the latter \enquote{affective view} \citep{kaminski2019Begriffe, coeckelbergh2012can}. 
The predictive view of trust is largely fed by rational choice theory approaches that think in terms of decision-making \citet{kaminski2019Begriffe}. Here, trust results from a rational choice of an actor who assess a given situation in terms of the probability of the negative and positive outcome. If it is more likely that the situation will result in a positive outcome for her, the rational actor will trust.

In contrast, the affective view highlights those cases in which trust is not based on any sort of probability assessment but rather something affective, related to a pre-reflexive attitude, even a precondition of acting within a social community \citep{coeckelbergh2012can}. Paradigmatic examples for this understanding of trust are intimate personal relationships, between friends, or parents and children. Remarkably, understanding trust in this way leads to the insight that trust is an underlying, non-thematic phenomenon which only shows itself in times of crisis. As long as you trust and the trusting relationship is working out, there is no need to talk about or even think about it---quite the opposite to the rational-choice assumptions \citep{baier1986trust}\cite[p. 29]{luhmann2000vertrauen}. 

We propose that both views of trust might be helpful for the further exploration of healthy distrust in AI. On the one hand, the idea of a rational agent is a normative prototype of ideal decision-making. It allows to ask who should decide, when, under which conditions, and for what reason to trust or distrust an AI system. If we furthermore assume that trusting AI would be the rational choice based on a probability assessment, we need to clarify the necessary conditions for making such a rational probability assessment for AI systems. (These assumptions are pretty much in line with understanding trustworthy AI as a task of requirement engineering and the strategy of providing certificates for trustworthy AI.) On the other hand, we may also follow the idea that trust is not (only) something that we actively choose, on rational calculation, but is rather something that is always a precondition for our doings and relationships we find ourselves in. The affective view offers two aspects worth debating: One, is the emotional attachment to AI beneficial for healthy distrust? Two, what does it mean to understand trust in AI as something that is already always grounded in a social web of trusted relationships? We might assume, in parallel to family members and friends, that healthy distrust will become harder to cultivate the more affective we are invested in our relationship to an AI system (see above). We might further assume that deliberatively making a choice about whether or not to distrust an AI is always based on trusting something else and others (see below).

Third, to further discuss (dis-)trust and AI, we propose to use Niklas Luhmann’s \citeyear{luhmann1990technology} distinction between risks and dangers to model different relationships to the AI: \enquote{Risk is an attribute of decision-making, while danger is a condition of life in general that cannot be avoided.} \cite[p. 225]{luhmann1990technology}. Most trust theories argue that taking a risk is a necessary condition for trusting because if there were no risk to take there would be no need to trust. However, following Luhmann, we might argue that taking a risk is always bound to a self-made decision; it holds an internal relation to decision-making. Someone who is not making the decision is not taking a risk. Danger, in contrast, is something that comes upon you, independent of your decision-making, a threat you are being subjected to by circumstances: 

\begin{quote}
    Risk is not simply the lack of safety, but rather the possible damage that may result from one’s own  decisions. The antonym of risk, then, would be danger as possible damage stemming from external sources.\cite[p. 223]{luhmann1990technology}
\end{quote}

Speaking of risks and dangers implies a specific relation to the future and the expectations of what might happen. The risk category is bound to an actor who, by making a certain decision, is willing to take a certain risk, e.g., a 16th century sailor crossing the Atlantic is taking the risk of being captured by pirates. Taking a risk in that understanding is bound to an epistemic and a control condition: if you were such a sailor and if you were unaware of pirates you couldn't take the risk; if you were a slave and forced to sail the ocean you are not taking a risk as it was not your decision to do so, but you find yourself subject to the danger of being captured by pirates. Risk is bound to knowledge and according technologies, as well as a certain amount of autonomy to act as a free actor. Here, it is important that Luhmann argued from a sociological perspective, i.e., taking a risk or being subject to danger is bound to societal conditions. To give a simple and mundane example that Luhmann is famous for: umbrellas \citep[p. 201]{baraldi2021unlocking}. Once umbrellas are invented in a society and are regularly available, everyone who is not taking an umbrella with them runs the risk of getting wet when it rains. Once you live in an umbrella-society, rain is not a danger beyond your control anymore, as you are practically free to decide to protect yourself from rain with umbrellas. Instead, becoming wet by rain constitutes a risk. This means that risks are created by technology and knowledge. For example, once a prenatal diagnosis is available on a societal level, becoming parents are forced to make a decision of whether to do such a diagnosis or not. Even if they would try to ignore this option, it would practically mean that they decide not to do it, and, hence, by having to make this decision, they have to take the risk of either knowing or not knowing whatever insights the prenatal diagnosis could have provided them. 

From Luhmann’s perspective, the concept of risk is a relational one, incorporating societal constellations and conditions; both on a collective level---what is known and technically doable---, and on an individual level---what is part of your scope of decision-making. There are many potentially harmful things people find themselves subjected to which lie outside their individual scope of decision-making \citep{luhmann2017risk}. This is largely true for social systems in general: members of a society are rather subjected to social systems than actively deciding to be a part of such systems---and, hereby, taking certain risks. For instance, if you happen to live near to Ukraine or France as a regular citizen, you are subjected to the danger of an potential accident in one of their nuclear power plants (e.g. in the course of a bombing raid). The political actors, however, who decide how to run a state’s energy system, are taking the risk of operating nuclear power plants---while those political actors, such as recent Germany, who decide against using nuclear power plants, take the risk of not using them, hence being dependent on other energy source etc.. In a historical perspective, we can see that it is a typical trait of modern societies to turn dangers into risks on a societal level, which brings along power issues because of the asymmetrical positions of being part of decision-making processes or not. 
\enquote{Risk is the hopefully avoidable causal link between decision and damage} \citep[p. 225]{luhmann1990technology}.

While danger is a reference to the future in which one has no practical relationship, in the sense that one cannot directly influence the occurrence of this danger through one’s behavior or decisions (leaving only the option to face dangers with confidence), the category of risk brings with it a different reference to the future; it connects future events with one’s own ability to act. If the future is decoupled from one’s own ability to act, it is the same for everyone involved. If an earthquake creates a danger, everyone is equally affected. However, if there is collective knowledge and technology to transform the dangers of an earthquake into risks, the situation changes and this possible future can be related to one’s own ability to act. This raises questions of power, as not everyone has the same decision-making capacity.  

Transferring these arguments on the situation of using AI, we differentiate between different users of AI. An AI practitioner takes a risk in using AI while the future brought about by the AI-assisted decision-making constitutes a danger for those affected by the decision-making. For those affected---the data subjects in the terminology of the GPDR---using AI becomes part of the wider social systems they find themselves subjected to anyway, and which becomes practically hard, if not impossible, to avoid. The situation is different for those who have agency in the associated decision-making process, that is, whether or not to use certain AI systems for certain domains at all, whether or not to regulate these systems or not, and how exactly to deploy these systems in a company, administration, etc., and deciding to follow a recommendation or not. Therefore, deploying AI practically means quite different things for us depending on our societal position. More to the point, for Luhmann, only taking a risk is linked to trust while danger is linked to confidence (in German: \enquote{Zuversicht}), that is, from his perspective, it is meaningless to expect data subjects to trust in AI. What they can do, as citizens, is to be critical about the social systems the AI is embedded in, e.g., as a student to question the quality of education in an institution where it has become normal to be completely assisted by AI tools in reading and writing anything. 


\section{Concepts of trust and distrust in psychology}

In the present paper, we attempt to conceptualize the idea of positive or beneficial distrust, something we would like to call \emph{healthy distrust} for the time being. In order to clarify what such healthy distrust could be, we first turn to psychology. To preview those results that will be important for answering our questions, psychological research has (i) demonstrated benefits of a distrust mindset and (ii) pointed out that trust and distrust are not mutually exclusive. It has, however, (iii) neither developed a systematic account of distrust (iv) nor can its ideas easily be adapted to attitudes towards AI output.

In the following, we will spell out the involvement of psychology with and its understanding of distrust and discuss how far it can get us in developing an appropriate understanding of distrust within explainable AI. Because research on distrust is still rather fragmentary, we will have to include conclusions from trust research.

Although the phenomenon of trust is, as mentioned above, pervasive in all aspects of modern life \citep{frevert2013Vertrauensfragen}, the topic has long been marginal in psychology \citep{clases2000Vertrauen}. This applies even more to distrust. Its study is scattered across different subdisciplines (among others developmental, organizational, social, and clinical psychology), often with different foci and methods, and, what is more, different basic conceptions of humans. For instance, game-theoretical approaches of behavioral economics see people as calculating and profit-maximizing, whereas psychoanalytical approaches focus on the fundamental conflictuality of mental life; and cognitive theories regard humans as information processors. Understandings of (dis)trust cannot be integrated across such different approaches. 

What is more, the definitions of (dis)trust change depending on the respective psychological approach: Trust and distrust may be understood as elements of decision processes, as something to be fostered as basics of a healthy personality, as stable personality traits, or transient states within individual information processing or temporary work groups.
It is not yet clear whether there are enough commonalities in the phenomena studied in different approaches to treat them as instances of the same trust or distrust concept, even though there are some similarities, for instance, the presence of risk, however abstract, the implication of vulnerability, the emergence of options for action, and a state between knowing and not knowing \citep{clases2000Vertrauen}.

Two early approaches to trust were those of psychoanalyst Erik Erikson and social learning theorist Julian Rotter. For Erikson, basic trust, which is developed in early childhood within the mother-infant relationship and remains for life, is a necessary element of a healthy personality \citep{erikson1950growth}. Mistrust may lead to undesirable personality development---although, in Erikson’s opinion, a certain amount of mistrust is necessary, too; see below. Erikson’s ideas had a significant impact on Western people’s view of themselves in their time and still resonate today in the popularity of attachment theories. However, as stable features, they have little substantive connection to the (dis)trust that is at stake when interacting with AI.

\citet[][p. 651, p. 653]{rotter1967new} defined interpersonal trust as the \enquote{expectancy held by an individual that the word, promise, verbal or written, statement of another individual group or person others can be relied upon}, more specifically, a \enquote{generalized expectancy} or \enquote{generalized attitude} learned from experience. While underscoring that his definition departs strongly from the definition of trust by Erikson, he nevertheless lists a variety of relations in which the ability to trust (or the lack thereof) is important: family, representatives of society, majority/minority, psychotherapy, and learning in general (the latter because it relies on the trust that texts are correct). Trust, thus, is an important aspect of social interactions. At first glance, Rotter’s definition seems to be applicable to the case of AI. However, he explicitly targets \textit{interpersonal} trust which may have different preconditions, workings, and consequences than human-AI trust. Secondly, he was particularly interested in generalized trust as a stable, learned trait. This also limits the application of his concept, as AI researchers would benefit more from ideas of how trust concerning a specific system or within a specific situation arises than from understanding stable trust. 

Both approaches, Erikson’s as well as Rotter’s are examples of the valorization of trust that Frevert observed in modern Western society, and they understand mistrust or distrust as the opposite of trust. Later psychological research has shown that this latter contrast is too narrow.


Much later research within psychology is more closely related to the AI context because it is in the applied domain, focusing antecedents and consequences of trust. Within basic research in developmental psychology, the development of trust and distrust in interlocutors by children has always played an important role. Also, distrust is a relevant topic in communication. Humans have to detect deception and misinformation \citep{cosmides1992tam}, and it seems that a mindset of suspicion can be easily caused by betrayals or signs of unreliability \citep{MarchandVonk2005BecomingSuspicious}. Further examples are work within organizational psychology which tries to understand and foster trust of leaders or within working groups, or in consumer psychology targeting brand trust, in clinical psychology on the role of trust in psychotherapy, studies on science communication with a focus on increasing the public’s trust in science, and, in the last decades, trust in automation, robots, online communication and, of course, AI.

In many of these studies, distrust was assumed to be the absence of or opposite of trust and, thus, of little interest in itself. As a research topic it became more interesting when empirical evidence emerged that it has some specific benefits. To give an example, a state of distrust activates cognitions incongruent with a message’s meaning \citep{SchulEtAl2004Encoding}. Such a distrust mindset does not seem to be beneficial in general, but individuals in such a state outperform trusting individuals in unusual environments, that is, in situations in which routine strategies are less helpful than non-routine strategies \citep{schul2008JEPS}. As \citep{mayermussweiler2011Suspicious} reported, distrust can also enhance creativity. This benefit was caused by increased cognitive flexibility and, thus, accords well with the results reported by \cite{SchulEtAl2004Encoding, schul2008JEPS}. 

One prominent definition of distrust has been put forward by \citet[][p. 439]{lewicki1998trust&distrust} in the context of organizational psychology. They define distrust \enquote{in terms of confident negative expectations regarding another’s conduct}, the latter summarized as \enquote{another’s words, actions, and decisions (what another says and does and how he or she makes decisions)}.  They continue to explain the confident negative expectations as \enquote{a fear of, a propensity to attribute sinister intentions to, and a desire to buffer oneself from the effects of another’s conduct.} High distrust is, therefore, in the definition of \citet[][p. 445]{lewicki1998trust&distrust}, characterized by fear, skepticism, cynicism, wariness, watchfulness, and vigilance. These features cover a broad spectrum in terms of evaluation and include the vigilance and watchfulness that may be helpful when dealing with potential fallacies of AI.

However, directly transferring the definition of \cite{lewicki1998trust&distrust} to the AI case would presuppose to see AI as a human-like agent to whom intentions are attributed. While humans certainly often and regularly anthropomorphize robots \citealp{deshpande2023anthropomorphizationaiopportunitiesrisks, placani2024anthropomorphism}, computers, AI and the like \citep{scorici2024anthropomorphization}, an understanding of healthy distrust should not presuppose anthropomorphization. Quite the opposite: As spelled out above, reasons for distrust in AI typically emerge in situations in which negative intentions are absent.

Another definition has been put forward by \citep[][p. 1293]{schul2008JEPS}: \enquote{Distrust is a ubiquitous psychological state that arises whenever we seem unable to take appearances, typically others’ declarations and behaviors, at face value. This usually occurs because we recognize that their interests conflict with ours. Somewhat less common are instances in which we are unaware of any explicit conflict of interest but nonetheless feel uneasy, sensing that the situation is not normal, that others may behave unpredictably or cause something unexpected to happen.} Here, the focus on others' intentions (deception) is, though present \cite[see also][]{SchulEtAl2004Encoding}, less prominent. Also, it seems important to us that the authors focus not only on explicit distrust, but also include an uneasy feeling.

Psychologists also identified trait distrust, that is a \enquote{relatively enduring belief that others are untrustworthy, exploitative, and self-serving to one’s own disadvantage} \cite[focusing on sociopolitically flavored distrust][p. 791]{ThielmannHilbig2023GeneralizedDispositionalDistrust}. \cite{ThielmannHilbig2023GeneralizedDispositionalDistrust} explicitly argue that such distrust is dangerous for societies because it is one of the causal preconditions of populistic beliefs and conspiracy mentality.

It is important to regard distrust and trust not as opposites and \cite{lewicki1998trust&distrust} seem to be the first to have clearly stated this. 
\cite{baselbrühl2023Misstrauen} summarized existing conceptualizations of distrust in the form of three models. In the first model, trust and distrust are mutually exclusive opposites. In addition to Luhmann’s sociological approach, Rotter’s generalized expectations, Erikson’s basic (mis)trust and game-theoretic approaches can be classified into this group.
The second model responds to the empirical finding that trust and distrust can coexist. It is a weaker version of the first model because it allows for a neutral zone in which neither trust nor distrust prevails, but rather an attitude of suspicion in which people search for information about the intentions of another person. For this purpose, they engage in attributions, which is another reason why this model is more attractive for cognitive psychologists.
Only the third model assumes trust and distrust to be two separate, yet related dimensions which can coexist. Agreeing with this model---which we do in principle---requires us, however, to define trust and distrust more precisely.

After repeatedly pointing out the value of a distrust mindset \cite[][p. 3]{mayo2024neither}, in the context of problems of misinformation, recently argued that neither a mindset of trust nor one of distrust is sufficient. The term mindset is broader than the notions of expectation or expectancy used by \cite{lewicki1998trust&distrust} as well as \cite{rotter1967new}. Mindsets are cognitive orientations and procedures applied to certain tasks; they can be activated by personality variables, learning, and situational cues. While trust mindsets facilitate social interactions, they also lead to specific biases such as the truth bias \citep{pantazi2018truthbias}, or the illusory truth effect \citep{UdryBarber2024IllusoryTruthReview}. Distrust mindsets (similarly related to personality and situation variables) render people more sensitive to disinformation and counter their inclination to trust by believing less and considering more alternatives, but also push them towards disbelieving correct information and considering alternatives to them. \cite[][p. 3]{mayo2024neither} therefore dubbed the distrust mindset a \enquote{double-edged sword}.

For dealing with misinformation, Mayo therefore proposed what she terms a \textit{Cartesian model},\footnote{As opposed to the more common \enquote{Spinozan} model. In the Spinozan model, trust or belief is primary and evaluative processes are secondary. Much psychological research implicitly adheres to this model by assuming that belief is automated and effortless, whereas disbelief is effortful, and prone to failure. The most compelling (but not the only) evidence for the Spinozan model is the truth bias \citep{pantazi2018truthbias, walter2020misinformation}.} which \enquote{posits that individuals can comprehend information without hastily labeling it as true or false} and \enquote{involving a pause before automatically accepting or rejecting to carefully consider the content}. The term Cartesian here refers to Cartesian doubt, a methodical approach used by the early modern philosopher René Descartes to first question any supposed knowledge and identify reasons for its falsity (with the goal to identify an undoubtable foundation of knowledge). 

This Cartesian model represents what Mayo calls an evaluative mindset in which individuals prioritize the \emph{accuracy} of the information they encounter and thereto hold back with both trust and distrust. Importantly, the Cartesian mindset does not imply to think more or less, but rather to pause. According to Mayo, it can be established by prioritizing accuracy.

There are empirical results that priming for accuracy---essentially cultivating this evaluative mindset---can indeed be beneficial when dealing with potential misinformation \citep{salovich2022evaluative}. Importantly, Mayo notes that the benefits of an evaluative mindset are contingent on prior knowledge, which limits its applicability. Individuals with little knowledge of a certain task---which might be exactly those cases in which many people seek AI advice---will not benefit and emphasis on accuracy may not be as relevant. 

Furthermore, while prioritizing accuracy is certainly important, it falls short of what we aim at, here. Healthy distrust may best be understood not as an individual mindset but as humans enacting their freedom and position themselves critically, not only sceptically, whereas understanding it as a mindset only might be an example of Western psychologization of societal issues \citep{adolffson2025Western}.

To sum up, psychology initially started with an appraisal of trust as predominantly positive and distrust as predominantly negative. This has been attenuated in recent years by findings that distrust has at least some cognitive benefits.
Psychology can provide AI research with some important distinctions such as the one between distrust as trait and as state. However, its distrust concept is closely linked to detection of dishonesty or malicious intentions in human-human interaction. It seems unlikely to us that the mechanisms that humans have developed for this can play a major role in interaction with AI. Mayo’s idea of a Cartesian mindset might be more suitable here, but it comes at the expense of the expertise that it requires, and it is an individual mindset which may fall short of our case.

Why do we suggest the term \enquote{distrust} at all, if no (psychological) definition of it matches our expectations? The main reason is that (X)AI studies almost exclusively focus on trust. They take it for granted that trust, and only trust, is desirable. That this is the case is probably due, in no small part, to the unreservedly high value placed on trust---as described by Frevert---that has characterized significant societies for several centuries. Such \enquote{discursive dominance} must be rhetorically countered with something striking, which---hopefully---is achieved by the pairing of the clearly negative word mistrust with the positive word healthy.\footnote{Note that our wording \enquote{healthy distrust} does not stem from \cite[][p. 450]{lewicki1998trust&distrust}’s statement that \enquote{that social structures appear most stable where there is a healthy dose of both trust and distrust}.}

\section{A critical view on healthy distrust}

Despite the minor role that distrust has played, historically, the notion of healthy distrust appears plausible for modern, Western societies. It conforms to the skeptical attitude that we relate to rational, autonomous citizens; \enquote{skeptical} not in the Cartesian sense to doubt everything, but rather following the Enlightenment idea that things should not just be believed (that is, grounded in religion) or commanded (that is, grounded in authority) but rather established by inquiry, arguments, research, etc. (that is, grounded in reason). In modern societies however, this general valuation of a skeptical attitude has become frustrated. As classical positions in the sociology of modernity have established, in modern societies, labor is highly distributed, including epistemic labor. Thus, it is necessary to trust the systems that produce insight (relating back to the notion of system trust by \cite{luhmann2000vertrauen} mentioned above). Furthermore, the complex practices and technologies that are necessary to provide reasons for knowledge, at the same time, open up ever more things we do not know and that need to be trusted, too \citep{beck1996wissen,luhmann1992oekologie}. However, Wittgenstein has argued that this inherent necessity of every human practice to trust something that cannot be questioned or doubted without this practice loosing its sense is only a problem for too high philosophical expectations \citep{wittgenstein1969certainty}. In contrast, for everyday actors this is just normal. Only in rare occasions do we doubt certainties we normally trust---which needs another form of acting that, in turn, needs to trust other things. Following Wittgenstein, we can doubt everything but not everything at the same time \citep[§115]{wittgenstein1969certainty}. 

The notion of healthy distrust is caught in this predicament particularly through the requirement of being \enquote{healthy}. In order to be \enquote{healthy}, distrust needs to be rooted in some kind of knowledge, reasoning, or at least intuition about the world. That, however, makes it necessary to trust something else, that is this particular relationship to the world.

While Wittgenstein ties this difference between trust and rare occasions of distrust to a rather abstract notion of \enquote{matter-of-course} or \enquote{everyday} acting, it is important to note that the \enquote{normality} that goes with trusting certain things unquestionably is tied to social structures, particularly power relations. Early on, \citet{zerilli1998doing} has used Wittgenstein’s theory to analyze feminist politics, arguing that they amount to a political struggle, and in that sense a power struggle, of what counts to be \enquote{ordinary}, rather than an epistemic struggle on \enquote{women} and their situation.

This analysis can be transferred to technology. Technology is one of the enablers of modern societies and made it particularly clear that individual, skeptical attitudes have their limits. In a sense, the very reason of technology is to enable users to do things they no longer need to know how to do themselves---which, however, presumes that users trust that technology works accordingly. This is the reason why Marxist theorists, widely conceived, have discussed technology as alienation or ideology \citep{Fuchs2021}. This usually leads to an attempt to re-establish the skeptical attitude towards technology itself, by means of protruding the ideology of usage and interfaces \citep{Galloway_2012}. Further strands of theory, even those quite hostile to Marxist thought, have tried to renew skepticism by considering the true successor of the enlightenment that person who understands technology in its material and functional essence \citep{Kittler_1992}. Both these strands of thought, however, tend to fix one element or one practice as given or even essential (be it social relations or patterns of transistors on a chip). This ignores whatever enables this givenness and at the same time enables the distrust of the interface or normal use. When everything can be doubted, even if not at the same time, this seems an arbitrary decision for one form of distrust and its enabling conditions---among many that are possible.

More recent theories of technology, informed by queer and feminist science and technology studies, provide an alternative to this search for a definitive basis of appropriate distrust. As Karen Barad argues, any \enquote{apparatus of observation} enables some observations while precluding others \citep{barad2007meeting}, without any possibility to say which one is the \enquote{best} or \enquote{right} one per se---at least, without further contextual information. This entails that, for example, the common use of a technology, say a piece of software, and its detailed analysis, such as reading its source code, are just different perspectives for different forms of usage---neither is better or closer to the truth per se \cite[chapter 3]{matzner2024algorithms}. They enable different practices, and both need to rely on certain things---the function of the program in the first case, disassemblers and code editors in the second, hardware and a lot more in both cases. However, this interplay of enabling and disabling---and, accordingly, of trust and distrust---is not just a physical or ontological feature of the world as it sometimes appears to be in Barad’s writing. 

As in the discussion above, this analysis needs to be amended with a perspective on power. What is possible in the usage of a technology, including what can be distrusted and what needs to be trusted, depends on contexts of power. For example, the many reported cases of racial bias in AI \citep{benjamin2019race} mean that white people can just use the technology. They do not even notice what they trust in doing so---among other things, the appropriate training of the AI model. Non-white persons, in contrast, experience that outright: they would need to rely on a model that keeps frustrating them and can only be distrusted. That is why, also, the common interplay of normal usage of technology and its breakdown is incomplete. Both normality and its breakdown are not just features of technology or certain forms of use but are rooted in power structures. In this case, normality is only for white people. At the same time, the failure of technology in the case of non-whites is (frustratingly) \enquote{normal}, given its racist bias. What is more, such experiences of being (technologically) excluded are (even more frustratingly) normal for people in non-hegemonic power positions \citep[p. 138]{matzner2024algorithms}. 

Thus, rather than based on normal and non-normal use, different forms of distrust (or lack thereof) need to be compared by what each of them needs to trust in turn. Each form of distrust needs to trust something else. Whether distrust is \enquote{healthy} or not, in turn, needs to be judged regarding what this distrust trusts, what drives and enables it. For healthy distrust of a technology, this needs to include some relation to what the technology does, or is meant to do. For example, when AI is used for a medical diagnosis, a healthy distrust of its results would at least require some experience with that diagnosis---a user trusts their own experience, which enables them to distrust an AI system's results that do not correspond to the experience.

This analysis has several consequences. For one, it shows that healthy distrust is not just something that can be started at any moment. It presumes a kind of alternative relation to whatever is done with technology that first needs to be established somehow, and second needs to be trusted. Such alternative relationships can be grounded in experience---either with the area of application, such as medical diagnosis in the above example, or with the technology. An experienced computer scientist, for example, might healthily distrust the result of a medical AI based on some artifacts in an image---without any medical knowledge. Experience could also stem from particular forms of training, or from reports about racist AI in the media. 

Second, these conditions for healthy distrust need to be seen in the context of power. As in the above example, for some users in some cases they might be almost certainly be given, for others the issue of distrust (or trust) does not even arise. They just do what they do. This extends common critiques of the Enlightenment-sceptical autonomous subject to the issue of trust and distrust. This subject is modeled after some specific subject position with particular socio-technical preconditions---and taking that as a model for all rational acting dissimulates both the specificity of this situatedness and its conditions \citep{atanasoski2019surrogate, Matzner_2019}.

This entails that there are socio-cultural resources for healthy distrust. Healthy distrust can arise out of detailed dealing with a technology and its functionality---this is the main focus of current debates on XAI---
and also data literacy or digital literacy more generally \citep{dignazio2020data}. Healthy distrust can also be based in social experiences. For example, noticing that an AI application continues forms of discrimination, exclusion, or exploitation, as is the case with racial biases in AI, can lead to healthy distrust just by the social effects of the technology and without any deeper inquiry into its functioning or technical features.

Taken together, these analyses mean that healthy distrust is a limited option for dealing with the problems of AI. Not everyone has the resources or social situation to acquire a trustworthy basis for healthy distrust in technology. Way beyond the discussions of AI as \enquote{black box}, this is an issue of complexity, but also intellectual property, corporate secrecy, time, and effort \citep{burrell2016}. Just having a vague intuition of what goes on in a usual smartphone is way beyond most users, including quite tech-savvy ones. This also entails the socio-cultural resources for distrust. Discriminatory effects of AI are centrally caused, yet individually felt, without too many resources other than a few NGOs and critical researchers to collect this distributed epistemic puzzle into a larger picture of platform business models and technological politics. Furthermore, this bigger picture has a hard time being noticed as long as the technology works flawlessly, in the sense of giving no reason to distrust for a sufficiently large, socially hegemonic group of users. 
That is, even if healthy distrust arises, it may be hard to act on it. Users may be forced to follow the results they distrust nevertheless, e.g., because workplace rules require it, or applications for welfare require it \citep{eubanks2017automating}. Since many AI applications come in corporate, fixed applications with strong rules regarding usage or even repair \citep{sajn_briefing_2022}, healthy distrust very often just leads to the alternative of not using a software at all or using it despite the distrust because the social or economic costs are too high. 

All of this has parallels to other attempts at finding individualist forms of dealing with technological problems \citep{Matzner_etal_2016, atanasoski2019surrogate}: even if epistemically feasible, individual possibilities for action may be too limited, and more institutional or regulatory approaches are needed. This does not mean that there is no place for healthy distrust, yet it may require a thorough regard of the social, cultural, and technological context to carve out the space where healthy distrust can lead to a more autonomous use of technology---and where the requirement to distrust or to build healthy distrust just becomes a burden too difficult to fulfill.

\section{Re-conceptualizing healthy distrust}

In our interdisciplinary explorations of trust and distrust, we have encountered a wide variety of notions of trust and distrust, as well as related concepts such as Luhmann’s confidence (\enquote{Zuversicht}), scepticism, epistemic vigilance, and more. As such, one may reasonable ask why yet another term, healthy distrust, is needed beyond the plethora of vocabulary that is already available. Further, we have also covered objections to healthy distrust, e.g., that it may undermine the functioning of society or that it may be an unhelpful concept in contexts where the prerequisite expertise is missing or where individuals have no option to act on their healthy distrust.

Still, we believe that existing notions do not sufficiently account for the observations we made in this article, namely that
\begin{itemize}
\item trustworthiness of an AI system does not imply that users do or should trust the system \citep{visser2025csr},
\item distrust isn't a property of a system, but rather related to a user’s stance toward a specific usage practice of an AI system in a specific social embedding (including power relations and norms),
\item distrust has affective components that are insufficiently covered by rational choice models, epistemic vigilance, or skepticism, and even by much of psychological research
\item distrust may be justified by the impulse to assert one’s autonomy in opposition to automation---even if the automation is institutionally enforced and beyond one’s control,
\item distrust may not only be justified but normatively appropriate---perhaps, healthy distrust is even a virtue.
\end{itemize}

We propose a new notion of \emph{healthy distrust} as a (preliminary) attempt to account for these observations. With \emph{distrust} we refer to a partially rational and partially affective, careful or negative stance when being confronted with a certain usage practice for AI systems in a specific socio-technical context, i.e., the intuition that something is not right about this usage practice. With the attribute \enquote{healthy} we wish to express a normative component: that distrust can be an appropriate, desirable state, despite the historic, negative connotation \citep{kaestner_etal2021trust}.

Some notion along the lines of healthy distrust is needed not only to close a conceptual gap between other (related) concepts, but also for practical reasons. In contemporary politics and economy, we observe that AI systems are often element of oppressive or exploitative structures and practices; and such practices trigger justified unease, hesitancy, opposition, and resistance that is insufficiently covered by prior work \citep{earnhart2024regulation} and for which \emph{healthy distrust} may be an appropriate descriptor. Establishing this descriptor may help to make this empiric phenomenon explicit and serve as a starting point for further empiric research, e.g., by operationalizing healthy distrust, and investigating its relationship to other constructs and behaviors.

In terms of education, healthy distrust may serve as a useful complement to the notion of AI literacy \citep{long2020what}, which centers the knowledge and skills required to use AI well, whereas healthy distrust focuses on the knowledge and skills required to recognize questionable usage practices. Indeed, it may well be that healthy distrust is a necessary pre-requisite for meaningful human oversight and human-in-the-loop concepts \citep{sterz2024quest}: Human overseers at least need to anticipate the possibility of failures, i.e., have some level of distrust, to intervene with an AI system.

Finally, from a normative perspective, any notion of meaningful informed consent towards AI usage practices hinges on humans engaging critically with the respective usage practice in the respective social context \citep{edenberg2020troubleshooting}. Healthy distrust may be a necessary, or at least helpful, part of such a critical stance.

Still, we acknowledge reasonable objections to healthy distrust. For one, the conceptual overlap to other terms is substantial, although we believe that a gap remains that healthy distrust can fill. Second, healthy distrust may lead to under-utilization of AI systems and, therefore, prevent utility gains that would result from more AI use. This objection appears particularly strong given that we include pre-rational hesitancy and reluctance in healthy distrust. Put bluntly: How long should an institution wait for everyone to get over their healthy distrust before, finally, an AI system can be used? Trust, after all, should be the default case to ensure that society can function and disrupting the functioning of society certainly requires strong reasons.

However, we believe that, for many contemporary AI usage practices, the delay imposed by healthy distrust is warranted. For one, as outlined before, the past has shown numerous cases of algorithmic discrimination, data theft, surveillance, and many other questionable practices and structures related to AI systems. But more broadly, AI hype confronts us regularly with new proposals for questionable AI usage practices in ever novel contexts, justified by vague pointers to the supposed general intelligence capabilities of language models or other AI systems \citep{mitchell2024debates}. Confronted with this, healthy distrust can be part of re-affirming human autonomy and taking the necessary time to deliberate newly proposed AI usage practices in their respective social context, let humans acquire the necessary knowledge and skills to engage meaningfully with the practice, and finally make an an informed decision about it. As such, healthy distrust should not be seen as in opposition to trust and reliance, but as necessary to establish preconditions for meaningful trust and AI usage practices that respect human autonomy.

\section*{Acknowledgement}

We gratefully acknowledge funding by the German Research Foundation (Deutsche Forschungsgemeinschaft, DFG): TRR 318/1 2021 ‐ 438445824.

\bibliographystyle{plainnat}
\bibliography{references}

\begin{thebibliography}{78}
\providecommand{\natexlab}[1]{#1}
\providecommand{\url}[1]{\texttt{#1}}
\expandafter\ifx\csname urlstyle\endcsname\relax
  \providecommand{\doi}[1]{doi: #1}\else
  \providecommand{\doi}{doi: \begingroup \urlstyle{rm}\Url}\fi

\bibitem[Adolfsson and Finyiza(2024)]{adolffson2025Western}
Johanna~Sofia Adolfsson and Gertrude Finyiza.
\newblock The western psychologization of global development: A cultural and
  decolonial approach.
\newblock \emph{Theory \& Psychology}, 34\penalty0 (6):\penalty0 713--735,
  2024.
\newblock \doi{10.1177/09593543241279122}.

\bibitem[Angwin et~al.(2016)Angwin, Larson, Mattu, and Kirchner]{Angwin2016}
Julia Angwin, Jeff Larson, Surya Mattu, and Lauren Kirchner.
\newblock Machine bias.
\newblock \emph{Pro Publica}, 2016.
\newblock URL
  \url{https://www.propublica.org/article/machine-bias-risk-assessments-in-criminal-sentencing}.

\bibitem[Atanasoski and Vora(2019)]{atanasoski2019surrogate}
Neda Atanasoski and Kalindi Vora.
\newblock \emph{Surrogate Humanity}.
\newblock Duke University Press, Durham, NC, USA, 2019.

\bibitem[Baier(1986)]{baier1986trust}
Annette Baier.
\newblock Trust and antitrust.
\newblock \emph{Ethics}, 96\penalty0 (2):\penalty0 231--260, 1986.
\newblock \doi{10.1086/292745}.

\bibitem[Barad(2007)]{barad2007meeting}
Karen Barad.
\newblock \emph{Meeting the universe halfway}.
\newblock Duke University Press, Durham, NC, USA, 2007.

\bibitem[Baraldi et~al.(2021)Baraldi, Corsi, and
  Esposito]{baraldi2021unlocking}
Claudio Baraldi, Giancarlo Corsi, and Elena Esposito.
\newblock \emph{Unlocking Luhmann; Luhmann in Glossario. I concetti
  fondamentali della teoria: A Keyword Introduction to Systems Theory}.
\newblock Bielefeld University Press, Bielefeld, Germany, 2021.
\newblock URL \url{https://www.transcript-verlag.de/978-3-8376-5674-9}.

\bibitem[Barocas et~al.(2023)Barocas, Hardt, and Narayanan]{Barocas2023}
Solon Barocas, Moritz Hardt, and Arvind Narayanan.
\newblock When is automated decision making legitimate?
\newblock In \emph{Fairness and Machine Learning: Limitations and
  Opportunities}. The MIT Press, Cambridge, MA, USA, 2023.
\newblock URL \url{https://fairmlbook.org/legitimacy.html}.

\bibitem[Basel and Br{\"u}hl(2023)]{baselbrühl2023Misstrauen}
J{\"o}rn Basel and Rolf Br{\"u}hl.
\newblock Misstrauen. eine interdisziplin{\"a}re bestandsaufnahme.
\newblock In J{\"o}rn Basel and Philipp Henrizi, editors, \emph{Psychologie von
  Risiko und Vertrauen: Wahrnehmung, Verhalten und Kommunikation}, pages
  271--301. Springer, Berlin/Heidelberg, Germany, 2023.
\newblock \doi{10.1007/978-3-662-65575-7_11}.

\bibitem[Beck(1996)]{beck1996wissen}
Ulrich Beck.
\newblock Wissen oder nicht-wissen? zwei perspektiven reflexiver
  modernisierung.
\newblock In Ulrich Beck, Anthony Giddens, and Scott Lash, editors,
  \emph{Reflexive Modernisierung}, pages 286--315. Suhrkamp, Frankfurt a.M.,
  Germany, 1996.

\bibitem[Benjamin(2019)]{benjamin2019race}
Ruha Benjamin.
\newblock \emph{Race After Technology: Abolitionist Tools for the New Jim
  Code}.
\newblock Polity, Cambridge, UK, 2019.

\bibitem[Birhane et~al.(2022)Birhane, Kalluri, Card, Agnew, Dotan, and
  Bao]{Birhane2022}
Abeba Birhane, Pratyusha Kalluri, Dallas Card, William Agnew, Ravit Dotan, and
  Michelle Bao.
\newblock The values encoded in machine learning research.
\newblock In \emph{Proceedings of the 2022 ACM Conference on Fairness,
  Accountability, and Transparency (FAccT)}, page 173–184, 2022.
\newblock \doi{10.1145/3531146.3533083}.

\bibitem[Buolamwini and Gebru(2018)]{Buolamwini2018}
Joy Buolamwini and Timnit Gebru.
\newblock Gender shades: Intersectional accuracy disparities in commercial
  gender classification.
\newblock In \emph{Proceedings of the 1st Conference on Fairness,
  Accountability and Transparency (FAccT)}, pages 77--91, 2018.
\newblock URL \url{https://proceedings.mlr.press/v81/buolamwini18a.html}.

\bibitem[Burrell(2016)]{burrell2016}
Jenna Burrell.
\newblock How the machine ‘thinks’: Understanding opacity in machine
  learning algorithms.
\newblock \emph{Big Data \& Society}, 3\penalty0 (1):\penalty0
  2053951715622512, 2016.
\newblock \doi{10.1177/2053951715622512}.

\bibitem[Buschmeier et~al.(2023)Buschmeier, Buhl, Kern, Grimminger, Beierling,
  Fisher, Groß, Horwath, Klowait, Lazarov, Lenke, Lohmer, Rohlfing, Scharlau,
  Singh, Terfloth, Vollmer, Wang, Wilmes, and Wrede]{Buschmeier2023}
Hendrik Buschmeier, Heike~M. Buhl, Friederike Kern, Angela Grimminger, Helen
  Beierling, Josephine Fisher, André Groß, Ilona Horwath, Nils Klowait,
  Stefan Lazarov, Michael Lenke, Vivien Lohmer, Katharina Rohlfing, Ingrid
  Scharlau, Amit Singh, Lutz Terfloth, Anna-Lisa Vollmer, Yu~Wang, Annedore
  Wilmes, and Britta Wrede.
\newblock Forms of understanding of {XAI}-explanations.
\newblock \emph{arXiv}, \penalty0 (2311.08760), 2023.
\newblock URL \url{https://arxiv.org/abs/2311.08760}.

\bibitem[Clases and Wehner(2000)]{clases2000Vertrauen}
Christoph Clases and Theo Wehner.
\newblock Vertrauen, 2000.
\newblock URL
  \url{https://www.spektrum.de/lexikon/psychologie/vertrauen/16374}.

\bibitem[Coeckelbergh(2012)]{coeckelbergh2012can}
Mark Coeckelbergh.
\newblock Can we trust robots?
\newblock \emph{Ethics and information technology}, 14:\penalty0 53--60, 2012.
\newblock \doi{10.1007/s10676-011-9279-1}.

\bibitem[Cosmides and Tooby(1992)]{cosmides1992tam}
Leda Cosmides and John Tooby.
\newblock Cognitive adaptations for social exchange.
\newblock In Jerome~H. Barkow, Leda Cosmides, and John Tooby, editors,
  \emph{The Adapted Mind: Evolutionary Psychology and the Generation of
  Culture}, pages 163--227. Oxford University Press, Oxford, UK, 1992.

\bibitem[Deshpande et~al.(2023)Deshpande, Rajpurohit, Narasimhan, and
  Kalyan]{deshpande2023anthropomorphizationaiopportunitiesrisks}
Ameet Deshpande, Tanmay Rajpurohit, Karthik Narasimhan, and Ashwin Kalyan.
\newblock Anthropomorphization of ai: Opportunities and risks, 2023.
\newblock URL \url{https://arxiv.org/abs/2305.14784}.

\bibitem[D'Ignazio and Klein(2020)]{dignazio2020data}
Catherine D'Ignazio and Lauren~F. Klein.
\newblock \emph{Data Feminism}.
\newblock The MIT Press, Cambridge, MA, USA, 2020.

\bibitem[Earnhart(2024)]{earnhart2024regulation}
Larry Earnhart.
\newblock Regulation of tech forces-technological backlash: The luddite
  perspective.
\newblock In Abeba~N. Turi and Pooja Lekhi, editors, \emph{Innovation,
  Sustainability, and Technological Megatrends in the Face of Uncertainties:
  Core Developments and Solutions}, pages 179--200. Springer Nature
  Switzerland, Cham, Switzerland, 2024.
\newblock \doi{10.1007/978-3-031-46189-7_12}.

\bibitem[Edenberg and Jones(2020)]{edenberg2020troubleshooting}
Elizabeth Edenberg and Meg~Leta Jones.
\newblock Troubleshooting ai and consent.
\newblock In Markus~Dirk Dubber, Frank Pasquale, and Sunit Das, editors,
  \emph{The Oxford Handbook of Ethics of AI}. Oxford University Press, Oxford,
  UK, 2020.
\newblock \doi{10.1093/oxfordhb/9780190067397.013.23}.

\bibitem[Erikson(1950)]{erikson1950growth}
Erik~H Erikson.
\newblock Growth and crises of the "healthy personality".
\newblock In M.~J.~E. Senn, editor, \emph{Proceedings of the Symposium on the
  healthy personality}, pages 91--146. Josiah Macy, Jr. Foundation, 1950.

\bibitem[Eubanks(2017)]{eubanks2017automating}
Virginia Eubanks.
\newblock \emph{Automating Inequality}.
\newblock St. Martin's Press, New York, NY, USA, 2017.

\bibitem[Freiman(2023)]{freiman2023making}
Ori Freiman.
\newblock Making sense of the conceptual nonsense ‘{trustworthy AI}’.
\newblock \emph{AI and Ethics}, 3\penalty0 (4):\penalty0 1351--1360, 2023.
\newblock \doi{10.1007/s43681-022-00241-w}.

\bibitem[Frevert(2013)]{frevert2013Vertrauensfragen}
Ute Frevert.
\newblock \emph{Vertrauensfragen: Eine Obsession der Moderne}.
\newblock UTB : 2185. C. H. Beck, Munich, Germany, 2013.
\newblock ISBN 9783406656095.

\bibitem[Fuchs(2021)]{Fuchs2021}
Christian Fuchs.
\newblock Foundations of communication/media/digital (in)justice.
\newblock \emph{Journal of Media Ethics}, 36\penalty0 (4):\penalty0 186--201,
  2021.
\newblock \doi{10.1080/23736992.2021.1964968}.

\bibitem[Galloway(2012)]{Galloway_2012}
Alexander~R. Galloway.
\newblock \emph{The interface effect}.
\newblock Polity, Cambridge, UK; Malden, MA, USA, 2012.
\newblock ISBN 978-0-7456-6252-7.

\bibitem[Goddard et~al.(2011)Goddard, Roudsari, and Wyatt]{Goddard2011}
Kate Goddard, Abdul Roudsari, and Jeremy~C Wyatt.
\newblock Automation bias: a systematic review of frequency, effect mediators,
  and mitigators.
\newblock \emph{Journal of the American Medical Informatics Association},
  19\penalty0 (1):\penalty0 121--127, 06 2011.
\newblock \doi{10.1136/amiajnl-2011-000089}.

\bibitem[Goldberg(2020)]{goldberg2020trust}
Sanford~C Goldberg.
\newblock Trust and reliance 1.
\newblock \emph{The Routledge handbook of trust and philosophy}, pages 97--108,
  2020.

\bibitem[Grodzinsky et~al.(2020)Grodzinsky, Miller, and
  Wolf]{grodzinsky2020trust}
Frances Grodzinsky, Keith Miller, and Marty~J Wolf.
\newblock Trust in artificial agents.
\newblock In \emph{The Routledge handbook of trust and philosophy}, pages
  298--312. Routledge, Milton Park, UK, 2020.

\bibitem[Hasan(2024)]{Hasan2024-HASWYA-2}
Ali Hasan.
\newblock Are you anthropomorphizing {AI}?, 2024.
\newblock URL
  \url{https://blog.apaonline.org/2024/08/20/are-you-anthropomorphizing-ai-2/}.

\bibitem[Hoffman et~al.(2023)Hoffman, Mueller, Klein, and
  Litman]{hoffman2023measures}
Robert~R. Hoffman, Shane~T. Mueller, Gary Klein, and Jordan Litman.
\newblock Measures for explainable {AI}: Explanation goodness, user
  satisfaction, mental models, curiosity, trust, and human-{AI} performance.
\newblock \emph{Frontiers in Computer Science}, Volume 5 - 2023, 2023.
\newblock \doi{10.3389/fcomp.2023.1096257}.

\bibitem[Huang et~al.(2024)Huang, Sun, Wang, Wu, Zhang, Li, Gao, Huang, Lyu,
  Zhang, Li, Liu, Liu, Wang, Zhang, Vidgen, Kailkhura, Xiong, Xiao, Li, Xing,
  Huang, Liu, Ji, Wang, Zhang, Yao, Kellis, Zitnik, Jiang, Bansal, Zou, Pei,
  Liu, Gao, Han, Zhao, Tang, Wang, Vanschoren, Mitchell, Shu, Xu, Chang, He,
  Huang, Backes, Gong, Yu, Chen, Gu, Xu, Ying, Ji, Jana, Chen, Liu, Zhou, Wang,
  Li, Zhang, Wang, Xie, Chen, Wang, Liu, Ye, Cao, Chen, and Zhao]{Huang2024}
Yue Huang, Lichao Sun, Haoran Wang, Siyuan Wu, Qihui Zhang, Yuan Li, Chujie
  Gao, Yixin Huang, Wenhan Lyu, Yixuan Zhang, Xiner Li, Zhengliang Liu, Yixin
  Liu, Yijue Wang, Zhikun Zhang, Bertie Vidgen, Bhavya Kailkhura, Caiming
  Xiong, Chaowei Xiao, Chunyuan Li, Eric Xing, Furong Huang, Hao Liu, Heng Ji,
  Hongyi Wang, Huan Zhang, Huaxiu Yao, Manolis Kellis, Marinka Zitnik, Meng
  Jiang, Mohit Bansal, James Zou, Jian Pei, Jian Liu, Jianfeng Gao, Jiawei Han,
  Jieyu Zhao, Jiliang Tang, Jindong Wang, Joaquin Vanschoren, John Mitchell,
  Kai Shu, Kaidi Xu, Kai-Wei Chang, Lifang He, Lifu Huang, Michael Backes,
  Neil~Zhenqiang Gong, Philip~S. Yu, Pin-Yu Chen, Quanquan Gu, Ran Xu, Rex
  Ying, Shuiwang Ji, Suman Jana, Tianlong Chen, Tianming Liu, Tianyi Zhou,
  William Wang, Xiang Li, Xiangliang Zhang, Xiao Wang, Xing Xie, Xun Chen, Xuyu
  Wang, Yan Liu, Yanfang Ye, Yinzhi Cao, Yong Chen, and Yue Zhao.
\newblock {TrustLLM}: Trustworthiness in large language models.
\newblock \emph{arXiv}, \penalty0 (2401.05561), 2024.
\newblock URL \url{https://arxiv.org/abs/2401.05561}.

\bibitem[Ilievski et~al.(2024)Ilievski, Hammer, van Harmelen, Paassen,
  Saralajew, Schmid, Biehl, Bolognesi, Dong, Gashteovski, Hitzler, Marra,
  Minervini, Mundt, Ngomo, Oltramari, Pasi, Saribatur, Serafini, Shawe-Taylor,
  Shwartz, Skitalinskaya, Stachl, van~de Ven, and Villmann]{Ilievski2024}
Filip Ilievski, Barbara Hammer, Frank van Harmelen, Benjamin Paassen, Sascha
  Saralajew, Ute Schmid, Michael Biehl, Marianna Bolognesi, Xin~Luna Dong,
  Kiril Gashteovski, Pascal Hitzler, Giuseppe Marra, Pasquale Minervini, Martin
  Mundt, Axel-Cyrille~Ngonga Ngomo, Alessandro Oltramari, Gabriella Pasi,
  Zeynep~G. Saribatur, Luciano Serafini, John Shawe-Taylor, Vered Shwartz,
  Gabriella Skitalinskaya, Clemens Stachl, Gido~M. van~de Ven, and Thomas
  Villmann.
\newblock Aligning generalisation between humans and machines.
\newblock \emph{arXiv}, \penalty0 (2411.15626), 2024.
\newblock URL \url{https://arxiv.org/abs/2411.15626}.

\bibitem[Ji et~al.(2024)Ji, Qiu, Chen, Zhang, Lou, Wang, Duan, He, Zhou, Zhang,
  Zeng, Ng, Dai, Pan, O'Gara, Lei, Xu, Tse, Fu, McAleer, Yang, Wang, Zhu, Guo,
  and Gao]{Ji2024}
Jiaming Ji, Tianyi Qiu, Boyuan Chen, Borong Zhang, Hantao Lou, Kaile Wang,
  Yawen Duan, Zhonghao He, Jiayi Zhou, Zhaowei Zhang, Fanzhi Zeng, Kwan~Yee Ng,
  Juntao Dai, Xuehai Pan, Aidan O'Gara, Yingshan Lei, Hua Xu, Brian Tse, Jie
  Fu, Stephen McAleer, Yaodong Yang, Yizhou Wang, Song-Chun Zhu, Yike Guo, and
  Wen Gao.
\newblock {AI} alignment: A comprehensive survey.
\newblock \emph{arXiv}, \penalty0 (2310.19852), 2024.
\newblock URL \url{https://arxiv.org/abs/2310.19852}.

\bibitem[Kaminski(2019)]{kaminski2019Begriffe}
Andreas Kaminski.
\newblock {Begriffe in Modellen: Die Modellierung von Vertrauen in
  Computersimulation und maschinellem Lernen im Spiegel der Theoriegeschichte
  von Vertrauen}.
\newblock In Nicole~J. Saam, Michael Resch, and Andreas Kaminski, editors,
  \emph{Simulieren und Entscheiden: ozialwissenschaftliche Simulationen und die
  Soziologie der Simulation}, pages 173--197. Springer VS, Wiesbaden, 2019.
\newblock ISBN 978-3-658-26042-2.
\newblock \doi{10.1007/978-3-658-26042-2_7}.

\bibitem[Kastner et~al.(2021)Kastner, Langer, Lazar, Schomacker, Speith, and
  Sterz]{kaestner_etal2021trust}
Lena Kastner, Markus Langer, Veronika Lazar, Astrid Schomacker, Timo Speith,
  and Sarah Sterz.
\newblock On the relation of trust and explainability: Why to engineer for
  trustworthiness.
\newblock In \emph{Proceedings of the 29th International Requirements
  Engineering Conference Workshops (REW)}, pages 169--175, Los Alamitos, CA,
  USA, September 2021. IEEE Computer Society.
\newblock \doi{10.1109/REW53955.2021.00031}.

\bibitem[Kaur et~al.(2022)Kaur, Uslu, Rittichier, and Durresi]{Kaur2022}
Davinder Kaur, Suleyman Uslu, Kaley~J. Rittichier, and Arjan Durresi.
\newblock Trustworthy artificial intelligence: A review.
\newblock \emph{ACM Computing Surveys}, 55\penalty0 (2), January 2022.
\newblock \doi{10.1145/3491209}.

\bibitem[Kittler(1992)]{Kittler_1992}
Friedrich Kittler.
\newblock There is no software.
\newblock \emph{Stanford Literature Review}, 9\penalty0 (1):\penalty0 81–90,
  1992.

\bibitem[Körber et~al.(2018)Körber, Baseler, and Bengler]{Koerber2018trust}
Moritz Körber, Eva Baseler, and Klaus Bengler.
\newblock Introduction matters: Manipulating trust in automation and reliance
  in automated driving.
\newblock \emph{Applied Ergonomics}, 66\penalty0 (Munich):\penalty0 18--31, jan
  2018.
\newblock \doi{10.1016/j.apergo.2017.07.006}.

\bibitem[Lahno(2020)]{lahno2020trust}
Bernd Lahno.
\newblock Trust and emotion.
\newblock In \emph{The Routledge handbook of trust and philosophy}, pages
  147--159. Routledge, Milton Park, UK, 2020.

\bibitem[Laux et~al.(2024)Laux, Wachter, and Mittelstadt]{Laux2024}
Johann Laux, Sandra Wachter, and Brent Mittelstadt.
\newblock Trustworthy artificial intelligence and the {European Union AI act}:
  On the conflation of trustworthiness and acceptability of risk.
\newblock \emph{Regulation \& Governance}, 18\penalty0 (1):\penalty0 3--32,
  2024.
\newblock \doi{https://doi.org/10.1111/rego.12512}.

\bibitem[Lewicki et~al.(1998)Lewicki, McAllister, and
  Bies]{lewicki1998trust&distrust}
Roy~J. Lewicki, Daniel~J. McAllister, and Robert~J. Bies.
\newblock Trust and distrust: New relationships and realities.
\newblock \emph{Academy of Management Review}, 23\penalty0 (3):\penalty0
  438--458, 1998.
\newblock ISSN 0363-7425.
\newblock \doi{10.5465/amr.1998.926620}.

\bibitem[Long and Magerko(2020)]{long2020what}
Duri Long and Brian Magerko.
\newblock What is {AI} literacy? competencies and design considerations.
\newblock In \emph{Proceedings of the 2020 Conference on Human Factors in
  Computing Systems (CHI)}, CHI '20, page 1–16, 2020.
\newblock \doi{10.1145/3313831.3376727}.

\bibitem[Luhmann(1990)]{luhmann1990technology}
Niklas Luhmann.
\newblock Technology, environment and social risk: a systems perspective.
\newblock \emph{Industrial Crisis Quarterly}, 4\penalty0 (3):\penalty0
  223--231, 1990.
\newblock URL \url{http://www.jstor.org/stable/26162777}.

\bibitem[Luhmann(1992)]{luhmann1992oekologie}
Niklas Luhmann.
\newblock \emph{{\"O}kologie des Nichtwissens}, pages 149--220.
\newblock VS Verlag f{\"u}r Sozialwissenschaften, Wiesbaden, 1992.
\newblock \doi{10.1007/978-3-322-93617-2_5}.

\bibitem[Luhmann(2000)]{luhmann2000vertrauen}
Niklas Luhmann.
\newblock \emph{Vertrauen}.
\newblock Lucius \& Lucius Verlagsgesellschaft mbH, Stuttgart, Germany, 4th
  edition, 2000.

\bibitem[Luhmann(2017)]{luhmann2017risk}
Niklas Luhmann.
\newblock \emph{Risk: a sociological theory}.
\newblock Routledge, Milton Park, UK, 2017.

\bibitem[Marchand and Vonk(2005)]{MarchandVonk2005BecomingSuspicious}
Miquelle A.~G. Marchand and Roos Vonk.
\newblock The process of becoming suspicious of ulterior motives.
\newblock \emph{Social Cognition}, 23\penalty0 (3):\penalty0 242--256, 2005.
\newblock \doi{10.1521/soco.2005.23.3.242}.

\bibitem[Matzner(2019)]{Matzner_2019}
Tobias Matzner.
\newblock The human is dead – long live the algorithm! human-algorithmic
  ensembles and liberal subjectivity.
\newblock \emph{Theory, Culture and Society}, 36\penalty0 (2):\penalty0
  123–144, 2019.
\newblock ISSN 0263-2764.
\newblock \doi{10.1177/0263276418818877}.

\bibitem[Matzner(2024)]{matzner2024algorithms}
Tobias Matzner.
\newblock \emph{Algorithms}.
\newblock Routledge, Milton Park, UK, 2024.

\bibitem[Matzner et~al.(2016)Matzner, Masur, Ochs, and von
  Pape]{Matzner_etal_2016}
Tobias Matzner, Philipp~K Masur, Carsten Ochs, and Thilo von Pape.
\newblock Do-it-yourself data protection—empowerment or burden?
\newblock In Ronald~Leenes Gutwirth, Serge and Paul~De Hert, editors,
  \emph{Data Protection on the Move}, page 277–305. Springer, Dordrecht,
  Netherlands, 2016.
\newblock \doi{10.1007/978-94-017-7376-8_11}.

\bibitem[Mayer and Mussweiler(2011)]{mayermussweiler2011Suspicious}
Jennifer Mayer and Thomas Mussweiler.
\newblock Suspicious spirits, flexible minds: when distrust enhances
  creativity.
\newblock \emph{Journal of Personality and Social Psychology}, 101\penalty0
  (6):\penalty0 1262--1277, 2011.
\newblock \doi{10.1037/a0024407}.

\bibitem[Mayo(2024)]{mayo2024neither}
Ruth Mayo.
\newblock {Trust or distrust? Neither! The right mindset for confronting
  disinformation}.
\newblock \emph{Current Opinion in Psychology}, 56:\penalty0 101779, 2024.
\newblock \doi{https://doi.org/10.1016/j.copsyc.2023.101779}.

\bibitem[Mitchell(2024)]{mitchell2024debates}
Melanie Mitchell.
\newblock Debates on the nature of artificial general intelligence.
\newblock \emph{Science}, 383\penalty0 (6689):\penalty0 eado7069, 2024.
\newblock \doi{10.1126/science.ado7069}.

\bibitem[Mühlfried(2023)]{mühlfried2018intro}
Florian Mühlfried.
\newblock Introduction: Approximating mistrust.
\newblock In Florian Mühlfried, editor, \emph{Mistrust: Ethnographic
  Approximations}, pages 7--22. transcript, Bielefeld, 2023.
\newblock \doi{10.14361/9783839439234-001}.

\bibitem[Pantazi et~al.(2018)Pantazi, Kissine, and Klein]{pantazi2018truthbias}
Myrto Pantazi, Mikhail Kissine, and Olivier Klein.
\newblock The power of the truth bias: False information affects memory and
  judgment even in the absence of distraction.
\newblock \emph{Social Cognition}, 36\penalty0 (2):\penalty0 167--198, 2018.
\newblock \doi{https://doi.org/10.1521/soco.2018.36.2.167}.

\bibitem[Peters and Scharlau(2025)]{peters2025fip}
Tobias~M. Peters and Ingrid Scharlau.
\newblock Interacting with fallible {AI}: Is distrust helpful when receiving
  {AI} misclassifications?
\newblock \emph{Frontiers in Psychology}, 2025.
\newblock \doi{10.3389/fpsyg.2025.1574809}.

\bibitem[Placani(2024)]{placani2024anthropomorphism}
Adriana Placani.
\newblock Anthropomorphism in {AI}: hype and fallacy.
\newblock \emph{AI and Ethics}, 4\penalty0 (3):\penalty0 691--698, 2024.
\newblock \doi{10.1007/s43681-024-00419-4}.

\bibitem[Rotter(1967)]{rotter1967new}
Julian~B Rotter.
\newblock A new scale for the measurement of interpersonal trust.
\newblock \emph{Journal of Personality}, pages 651--665, 1967.
\newblock \doi{https://doi.org/10.1111/j.1467-6494.1967.tb01454.x}.

\bibitem[Salovich et~al.(2022)Salovich, Kirsch, and
  Rapp]{salovich2022evaluative}
Nikita~A. Salovich, Anya~M. Kirsch, and David~N. Rapp.
\newblock Evaluative mindsets can protect against the influence of false
  information.
\newblock \emph{Cognition}, 225:\penalty0 105121, 2022.
\newblock \doi{10.1016/j.cognition.2022.105121}.

\bibitem[Scharowski et~al.(2024)Scharowski, Perrig, Aeschbach, von Felten,
  Opwis, Wintersberger, and Br{\"u}hlmann]{scharowski2024trust}
Nicolas Scharowski, Sebastian~AC Perrig, Lena~Fanya Aeschbach, Nick von Felten,
  Klaus Opwis, Philipp Wintersberger, and Florian Br{\"u}hlmann.
\newblock To trust or distrust trust measures: Validating questionnaires for
  trust in {AI}.
\newblock \emph{arXiv preprint arXiv:2403.00582}, 2024.
\newblock \doi{10.48550/arXiv.2403.00582}.

\bibitem[Schlicker et~al.(2025)Schlicker, Baum, Uhde, Sterz, Hirsch, and
  Langer]{schlicker2025we}
Nadine Schlicker, Kevin Baum, Alarith Uhde, Sarah Sterz, Martin~C Hirsch, and
  Markus Langer.
\newblock How do we assess the trustworthiness of {AI}? introducing the
  trustworthiness assessment model ({TrAM}).
\newblock \emph{Computers in Human Behavior}, page 108671, 2025.

\bibitem[Schul et~al.(2004)Schul, Mayo, and Burnstein]{SchulEtAl2004Encoding}
Yaacov Schul, Ruth Mayo, and Eugene Burnstein.
\newblock Encoding under trust and distrust: The spontaneous activation of
  incongruent cognitions.
\newblock \emph{Journal of Personality and Social Psychology}, 86\penalty0
  (5):\penalty0 668–679, 2004.
\newblock \doi{https://doi.org/10.1037/0022-3514.86.5.668}.

\bibitem[Schul et~al.(2008)Schul, Mayo, and Burnstein]{schul2008JEPS}
Yaacov Schul, Ruth Mayo, and Eugene Burnstein.
\newblock The value of distrust.
\newblock \emph{Journal of Experimental Social Psychology}, 44\penalty0
  (5):\penalty0 1293--1302, 2008.
\newblock \doi{10.1016/j.jesp.2008.05.003}.

\bibitem[Scorici et~al.(2024)Scorici, Schultz, and
  Seele]{scorici2024anthropomorphization}
Gabriela Scorici, Mario~D Schultz, and Peter Seele.
\newblock Anthropomorphization and beyond: conceptualizing humanwashing of
  ai-enabled machines.
\newblock \emph{AI \& SOCIETY}, 39\penalty0 (2):\penalty0 789--795, 2024.
\newblock \doi{10.1007/s00146-022-01492-1}.

\bibitem[Simon(2020)]{simon2020routledge}
Judith Simon.
\newblock \emph{The Routledge handbook of trust and philosophy}.
\newblock Routledge, Milton Park, UK, 2020.

\bibitem[Smuha(2019)]{EU2019}
Nathalie Smuha.
\newblock Ethics guidelines for trustworthy {AI}.
\newblock Technical report, EU High-Level Expert Group on Artificial
  Intelligence, 2019.
\newblock URL
  \url{https://digital-strategy.ec.europa.eu/en/library/ethics-guidelines-trustworthy-ai}.

\bibitem[Smuha et~al.(2021)Smuha, Ahmed-Rengers, Harkens, Li, MacLaren,
  Piselli, and Yeung]{Smuha2021}
Nathalie~A Smuha, Emma Ahmed-Rengers, Adam Harkens, Wenlong Li, James MacLaren,
  Riccardo Piselli, and Karen Yeung.
\newblock How the {EU} can achieve legally trustworthy {AI}: a response to the
  {European Commission}'s proposal for an artificial intelligence act.
\newblock \emph{SSRN}, 2021.
\newblock \doi{10.2139/ssrn.3899991}.

\bibitem[Sterz et~al.(2024)Sterz, Baum, Biewer, Hermanns, Lauber-R\"{o}nsberg,
  Meinel, and Langer]{sterz2024quest}
Sarah Sterz, Kevin Baum, Sebastian Biewer, Holger Hermanns, Anne
  Lauber-R\"{o}nsberg, Philip Meinel, and Markus Langer.
\newblock On the quest for effectiveness in human oversight: Interdisciplinary
  perspectives.
\newblock In \emph{Proceedings of the ACM Conference on Fairness,
  Accountability, and Transparency (FAccT)}, page 2495–2507, 2024.
\newblock \doi{10.1145/3630106.3659051}.

\bibitem[Thielmann and
  Hilbig(2023)]{ThielmannHilbig2023GeneralizedDispositionalDistrust}
Isabel Thielmann and Benjamin~E. Hilbig.
\newblock Generalized dispositional distrust as the common core of populism and
  conspiracy mentality.
\newblock \emph{Political Psychology}, 44\penalty0 (4):\penalty0 789--805,
  2023.
\newblock \doi{https://doi.org/10.1111/pops.12886}.

\bibitem[Troshani et~al.(2021)Troshani, Hill, Sherman, and
  and]{Troshani2021trust}
Indrit Troshani, Sally~Rao Hill, Claire Sherman, and Damien~Arthur and.
\newblock Do we trust in ai? role of anthropomorphism and intelligence.
\newblock \emph{Journal of Computer Information Systems}, 61\penalty0
  (5):\penalty0 481--491, 2021.
\newblock \doi{10.1080/08874417.2020.1788473}.

\bibitem[Udry and Barber(2024)]{UdryBarber2024IllusoryTruthReview}
Jessica Udry and Sarah~J. Barber.
\newblock The illusory truth effect: A review of how repetition increases
  belief in misinformation.
\newblock \emph{Current Opinion in Psychology}, 56:\penalty0 101736, 2024.
\newblock \doi{10.1016/j.copsyc.2023.101736}.

\bibitem[Visser et~al.(2025)Visser, Peters, Scharlau, and
  Hammer]{visser2025csr}
Roel Visser, Tobias~M. Peters, Ingrid Scharlau, and Barbara Hammer.
\newblock Trust, distrust, and appropriate reliance in {(X)AI}: a survey of
  empirical evaluation of user trust.
\newblock \emph{Cognitive Systems Research}, 2025.
\newblock accepted.

\bibitem[Walter and Tukachinsky(2020)]{walter2020misinformation}
Nathan Walter and Riva Tukachinsky.
\newblock A meta-analytic examination of the continued influence of
  misinformation in the face of correction: How powerful is it, why does it
  happen, and how to stop it?
\newblock \emph{Communication Research}, 47\penalty0 (2):\penalty0 155--177,
  2020.
\newblock \doi{https://doi.org/10.1177/0093650219854600}.

\bibitem[Wittgenstein(1969)]{wittgenstein1969certainty}
Ludwig Wittgenstein.
\newblock \emph{On certainty}, volume 174.
\newblock Basil Blackwell, Oxford, UK, 1969.

\bibitem[Zerilli(1998)]{zerilli1998doing}
Linda M.~G. Zerilli.
\newblock Doing without knowing: Feminism's politics of the ordinary.
\newblock \emph{Political Theory}, 26\penalty0 (4):\penalty0 435--458, 1998.
\newblock \doi{10.1177/0090591798026004001}.

\bibitem[Šajn(2022)]{sajn_briefing_2022}
Nikolina Šajn.
\newblock Briefing {Right} to repair.
\newblock Technical Report PE 698.869, European Parliamentary Research Service,
  Brussels, 2022.
\newblock URL
  \url{https://www.europarl.europa.eu/thinktank/de/document/EPRS_BRI(2022)698869}.

\end{thebibliography}
\end{document}